\documentstyle[twoside,psfig]{article}

\def\simge{\mathrel{%
   \rlap{\raise 0.511ex \hbox{$>$}}{\lower 0.511ex \hbox{$\sim$}}}}
\def\simle{\mathrel{
   \rlap{\raise 0.511ex \hbox{$<$}}{\lower 0.511ex \hbox{$\sim$}}}}

\def\simeqq{\,\,{\raise 1.2ex\hbox{?}\llap{$\simeq$}}}

\def\+{\mathrel{
   \rlap{\raise -0.970ex \hbox{$\mathbf\widehat{}$}}
{\lower 0.511ex \hbox{$\mathbf\bigcirc$}}}}
\def\-{\,\,\,\,\,\mathrel{
   \llap{\raise -0.500ex \hbox{$\mathbf\bigcirc$}}
{\lower 0.990ex \hbox{$\mathbf\widehat{}$}}}}

\def\beq{\begin{eqnarray}}
\def\eeq{\end{eqnarray}}

\catcode`\@=11
\long\def\@makefntext#1{
\protect\noindent \hbox to 3.2pt {\hskip-.9pt
$^{{\eightrm\@thefnmark}}$\hfil}#1\hfill}		

\def\thefootnote{\fnsymbol{footnote}}
\def\@makefnmark{\hbox to 0pt{$^{\@thefnmark}$\hss}}	

\def\ps@myheadings{\let\@mkboth\@gobbletwo
\def\@oddhead{\hbox{}
\rightmark\hfil\eightrm\thepage}
\def\@oddfoot{}\def\@evenhead{\eightrm\thepage\hfil
\leftmark\hbox{}}\def\@evenfoot{}
\def\sectionmark##1{}\def\subsectionmark##1{}}

\oddsidemargin=\evensidemargin
\def\@citex[#1]#2{\if@filesw\immediate\write\@auxout
	{\string\citation{#2}}\fi
\def\@citea{}\@cite{\@for\@citeb:=#2\do
	{\@citea\def\@citea{,}\@ifundefined
	{b@\@citeb}{{\bf ?}\@warning
	{Citation `\@citeb' on page \thepage \space undefined}}
	{\csname b@\@citeb\endcsname}}}{#1}}
 \newif\if@cghi
\def\cite{\@cghitrue\@ifnextchar [{\@tempswatrue
	\@citex}{\@tempswafalse\@citex[]}}
\def\citelow{\@cghifalse\@ifnextchar [{\@tempswatrue
	\@citex}{\@tempswafalse\@citex[]}}
\def\@cite#1#2{{[{#1}]\if@tempswa\typeout           
	{WSPC warning: optional citation argument
	ignored: `#2'} \fi}}
\def\@refcitex[#1]#2{\if@filesw\immediate\write\@auxout
	{\string\citation{#2}}\fi
\def\@citea{}\@refcite{\@for\@citeb:=#2\do
	{\@citea\def\@citea{, }\@ifundefined
	{b@\@citeb}{{\bf ?}\@warning
	{Citation `\@citeb' on page \thepage \space undefined}}
	\hbox{\csname b@\@citeb\endcsname}}}{#1}}
 \def\@refcite#1#2{{[{#1}]\if@tempswa\typeout     
        {WSPC warning: optional citation argument
	ignored: `#2'} \fi}}
 \def\refcite{\@ifnextchar[{\@tempswatrue
	\@refcitex}{\@tempswafalse\@refcitex[]}}

\renewcommand{\thefootnote}{\fnsymbol{footnote}}

\newcounter{sectionc}\newcounter{subsectionc}\newcounter{subsubsectionc}
\renewcommand{\section}[1] {\vspace{12pt}\addtocounter{sectionc}{1}
\setcounter{subsectionc}{0}\setcounter{subsubsectionc}{0}\noindent
	{\tenbf\thesectionc. #1}\par\vspace{5pt}}
\renewcommand{\subsection}[1] {\vspace{12pt}\addtocounter{subsectionc}{1}
	\setcounter{subsubsectionc}{0}\noindent
	{\bf\thesectionc.\thesubsectionc. {\kern1pt \bfit #1}}\par\vspace{5pt}}
\renewcommand{\subsubsection}[1] {\vspace{12pt}\addtocounter{subsubsectionc}{1}
	\noindent{\tenrm\thesectionc.\thesubsectionc.\thesubsubsectionc.
	{\kern1pt \tenit #1}}\par\vspace{5pt}}
\newcommand{\nonumsection}[1] {\vspace{12pt}\noindent{\tenbf #1}
	\par\vspace{5pt}}

\newcounter{appendixc}
\newcounter{subappendixc}[appendixc]
\newcounter{subsubappendixc}[subappendixc]
\renewcommand{\thesubappendixc}{\Alph{appendixc}.\arabic{subappendixc}}
\renewcommand{\thesubsubappendixc}
	{\Alph{appendixc}.\arabic{subappendixc}.\arabic{subsubappendixc}}

\renewcommand{\appendix}[1] {\vspace{12pt}
        \refstepcounter{appendixc}
        \setcounter{figure}{0}
        \setcounter{table}{0}
        \setcounter{lemma}{0}
        \setcounter{theorem}{0}
        \setcounter{corollary}{0}
        \setcounter{definition}{0}
        \setcounter{equation}{0}
        \renewcommand{\thefigure}{\Alph{appendixc}.\arabic{figure}}
        \renewcommand{\thetable}{\Alph{appendixc}.\arabic{table}}
        \renewcommand{\theappendixc}{\Alph{appendixc}}
        \renewcommand{\thelemma}{\Alph{appendixc}.\arabic{lemma}}
        \renewcommand{\thetheorem}{\Alph{appendixc}.\arabic{theorem}}
        \renewcommand{\thedefinition}{\Alph{appendixc}.\arabic{definition}}
        \renewcommand{\thecorollary}{\Alph{appendixc}.\arabic{corollary}}
        \renewcommand{\theequation}{\Alph{appendixc}.\arabic{equation}}
        \noindent{\tenbf Appendix \theappendixc #1}\par\vspace{5pt}}
\newcommand{\subappendix}[1] {\vspace{12pt}
        \refstepcounter{subappendixc}
        \noindent{\bf Appendix \thesubappendixc. {\kern1pt \bfit #1}}
	\par\vspace{5pt}}
\newcommand{\subsubappendix}[1] {\vspace{12pt}
        \refstepcounter{subsubappendixc}
        \noindent{\rm Appendix \thesubsubappendixc. {\kern1pt \tenit #1}}
	\par\vspace{5pt}}

\newcommand{\textlineskip}{\baselineskip=13pt}
\newcommand{\smalllineskip}{\baselineskip=10pt}

\def\eightcirc{
\begin{picture}(0,0)
\put(4.4,1.8){\circle{6.5}}
\end{picture}}
\def\eightcopyright{\eightcirc\kern2.7pt\hbox{\eightrm c}}

\newcommand{\copyrightheading}[1]
	{\vspace*{-2.5cm}\smalllineskip{\flushleft
	{\footnotesize Modern Physics Letters A, #1}\\
	{\footnotesize $\eightcopyright$\, World Scientific Publishing
	 Company}\\
	 }}

\def\abstracts#1#2#3{{
	\centering{\begin{minipage}{4.5in}\footnotesize\baselineskip=10pt
	\parindent=0pt #1\par
	\parindent=15pt #2\par
	\parindent=15pt #3
	\end{minipage}}\par}}

\def\keywords#1{{
	\centering{\begin{minipage}{4.5in}\footnotesize\baselineskip=10pt
	{\footnotesize\it Keywords}\/: #1
	 \end{minipage}}\par}}

\newcommand{\bibit}{\nineit}
\newcommand{\bibbf}{\ninebf}
\renewenvironment{thebibliography}[1]
	{\frenchspacing
	 \ninerm\baselineskip=11pt
	 \begin{list}{\arabic{enumi}.}
        {\usecounter{enumi}\setlength{\parsep}{0pt}
	 \setlength{\leftmargin 12.7pt}{\rightmargin 0pt} 
         \setlength{\itemsep}{0pt} \settowidth
	{\labelwidth}{#1.}\sloppy}}{\end{list}}

\newcounter{itemlistc}
\newcounter{romanlistc}
\newcounter{alphlistc}
\newcounter{arabiclistc}

\newcommand{\fcaption}[1]{
        \refstepcounter{figure}
        \setbox\@tempboxa = \hbox{\footnotesize Fig.~\thefigure. #1}
        \ifdim \wd\@tempboxa > 5in
           {\begin{center}
        \parbox{5in}{\footnotesize\smalllineskip Fig.~\thefigure. #1}
            \end{center}}
        \else
             {\begin{center}
             {\footnotesize Fig.~\thefigure. #1}
              \end{center}}
        \fi}

\newcommand{\tcaption}[1]{
        \refstepcounter{table}
        \setbox\@tempboxa = \hbox{\footnotesize Table~\thetable. #1}
        \ifdim \wd\@tempboxa > 5in
           {\begin{center}
        \parbox{5in}{\footnotesize\smalllineskip Table~\thetable. #1}
            \end{center}}
        \else
             {\begin{center}
             {\footnotesize Table~\thetable. #1}
              \end{center}}
        \fi}

\def\@citex[#1]#2{\if@filesw\immediate\write\@auxout
	{\string\citation{#2}}\fi
\def\@citea{}\@cite{\@for\@citeb:=#2\do
	{\@citea\def\@citea{,}\@ifundefined
	{b@\@citeb}{{\bf ?}\@warning
	{Citation `\@citeb' on page \thepage \space undefined}}
	{\csname b@\@citeb\endcsname}}}{#1}}

\newif\if@cghi
\def\cite{\@cghitrue\@ifnextchar [{\@tempswatrue
	\@citex}{\@tempswafalse\@citex[]}}
\def\citelow{\@cghifalse\@ifnextchar [{\@tempswatrue
	\@citex}{\@tempswafalse\@citex[]}}
\def\@cite#1#2{{$\null^{#1}$\if@tempswa\typeout
	{IJCGA warning: optional citation argument
	ignored: `#2'} \fi}}

\def\pmb#1{\setbox0=\hbox{#1}
	\kern-.025em\copy0\kern-\wd0
	\kern.05em\copy0\kern-\wd0
	\kern-.025em\raise.0433em\box0}

\def\fnt#1#2{\footnotetext{\kern-.3em
	{$^{\mbox{\scriptsize #1}}$}{#2}}}

\def\fpage#1{\begingroup
\voffset=.3in
\thispagestyle{empty}\begin{table}[b]\centerline{\footnotesize #1}
	\end{table}\endgroup}

\def\runninghead#1#2{\pagestyle{myheadings}
\markboth{{\protect\footnotesize\it{\quad #1}}\hfill}
{\hfill{\protect\footnotesize\it{#2\quad}}}}
\font\tenrm=cmr10
\font\tenit=cmti10
\font\tenbf=cmbx10
\font\bfit=cmbxti10 at 10pt
\font\ninerm=cmr9
\font\nineit=cmti9
\font\ninebf=cmbx9
\font\eightrm=cmr8

\def\qed{\hbox{${\vcenter{\vbox{			
   \hrule height 0.4pt\hbox{\vrule width 0.4pt height 6pt
   \kern5pt\vrule width 0.4pt}\hrule height 0.4pt}}}$}}

\renewcommand{\thefootnote}{\fnsymbol{footnote}}	

\begin{document}
\setlength{\textheight}{7.7truein}  

\runninghead{D. V. Ahluwalia}{Ambiguity in source flux of
cosmic/astrophysical neutrinos $ldots$}

\normalsize\textlineskip
\thispagestyle{empty}
\setcounter{page}{1}

\copyrightheading{}			

\vspace*{0.88truein}

\fpage{1}

\centerline{\bf AMBIGUITY IN SOURCE FLUX OF  COSMIC/ASTROPHYSICAL}
\baselineskip=13pt
\centerline{\bf  NEUTRINOS:
EFFECTS OF BI-MAXIMAL MIXING}
\centerline{\bf AND QUANTUM-GRAVITY INDUCED DECOHERENCE
\footnote{Work
supported by Consejo Nacional
de Ciencia y Tecnolog\'ia (CONACyT) under Project \# 32067-E.}}

\vspace*{0.37truein}
\centerline{\footnotesize D. V. AHLUWALIA}
\baselineskip=12pt

\centerline{\footnotesize\it Theoretical Physics Group, ISGBG, Ap. Pos. C-600}
\baselineskip=12pt
\centerline{\footnotesize\it Escuela de F\'isica de la UAZ,
Zacatecas, ZAC 98068, M\'exico}
\baselineskip=12pt
\centerline{\footnotesize\it E-mail: ahluwalia@phases.reduaz.mx;
http://phases.reduaz.mx}
\vspace*{10pt}


\vspace*{0.21truein}
\abstracts{
For high energy cosmic neutrinos Athar, Je\'zabek, and Yasuda (AJY)
have recently shown that the existing data on neutrino oscillations
suggests that
cosmic neutrino flux at the AGN/GRB source,
$F(\nu_e):F(\nu_\mu):F(\nu_\tau)\approx 1:2:0$, oscillates to
$F(\nu_e):F(\nu_\mu):F(\nu_\tau)\approx 1:1:1$. These results can be
confirmed at AMANDA, Baikal, ANTARES and NESTOR, and other
neutrino detectors with a good flavor resolution. Here, we re-derive the
AJY result from quasi bi-maximal mixing, and show that
observation of $F(\nu_e):F(\nu_\mu):F(\nu_\tau)\approx 1:1:1$ does not
necessarily establish cosmic neutrino flux at the AGN/GRB source to be
$F(\nu_e):F(\nu_\mu):F(\nu_\tau)\approx 1:2:0$.
We also note that if the length scale for the  quantum-gravity
induced de-coherence for astrophysical neutrinos is of the order
of a Mpc, then independent of the MNS matrix,
the Liu-Hu-Ge (LHG) mechanism  would lead to
flux equalization for the  cosmic/astrophysical neutrinos.
}{}{}

\vspace*{10pt}


\setcounter{footnote}{0}
\renewcommand{\thefootnote}{\alph{footnote}}

\vspace*{1pt}\textlineskip	
\section{Introduction}	
\vspace*{-0.5pt}
\noindent
The solar neutrino anomaly, the LSND excess events, and the
Super-Kamiokande
data on atmospheric neutrinos, find their natural explanation in terms of
oscillations of neutrinos from one flavor to another.\cite{jb,LSND,SuperK}
The only experiment so far that provides a direct evidence of oscillation
from one flavor to another is the LSND experiment. However, the LSND
result is still debated by the KARMEN collaboration.\cite{k}
It is expected
to be settled by the dedicated Fermi Lab. experiments. Nevertheless, a strong
tentative evidence for neutrino oscillations seems  established. An indirect
hint for neutrino oscillations resides in the problem of obtaining
successful models of type-II supernova explosions.  Neutrino oscillations,
if confirmed to exist, can significantly aid these
explosions:\cite{grf96} in effect, neutrino oscillations, provide an indirect
energy transport mechanism due to flavor-dependence of relevant
neutrino cross sections.

In this {\em Letter\/} we shall neglect any possible CP violation in
neutrino oscillations. We shall adopt the standard three-flavor neutrino
scheme. In that framework one can accommodate any of the following two
sets of data: (a) Data on the atmospheric neutrinos
and solar neutrino anomaly, or (b)
Data on atmospheric neutrinos and LSND excess events. In the quasi bi-maximal
mixing, the angle $\theta$, see Eq.\ (\ref{bm}) below, can accommodate either
the LSND results or the solar anomaly, but not both.

In the context of this experimental setting, and the stated
theoretical framework, this {\em Letter\/} establishes the abstracted result.
The origin for the abstracted result lies in the observation that
the observed $L/E$ flatness of the electron-like event ratio in the
Super-Kamiokande atmospheric neutrino data strongly favors\cite{qbm1,qbm2}
a quasi bi-maximal mixing matrix (and in fact this is what drives the AJY
result). Here we show that quasi
bi-maximal mixing transforms
$F(\nu_e):F(\nu_\mu):F(\nu_\tau)\approx 1:a:2-a$ to
$F(\nu_e):F(\nu_\mu):F(\nu_\tau)\approx 1:1:1$.
Note, the latter flux neither carries an $a$ dependence, nor is
it affected by the angle $\theta$.
This robustness has the consequence that by studying the
departures from the
$F(\nu_e):F(\nu_\mu):F(\nu_\tau)\approx 1:1:1$
for the observed cosmic
high energy flux one may be able to explore new and interesting sources/physics
of high energy cosmic neutrinos. The data, however, may also be used to study
unitarity-preserving deformations of bi-maximality.

In the next section,  we summarize the AJY result under study.
Section 3 shows the quasi bi-maximal mixing as the physical
origin of flux equalization for AGN/GRBs, it then presents the theorem
advertised in the {\em Abstract\/}, and it ends by introducing a deformed
bi-maximal mixing and its affect on the flux equalization.
Section 4 discusses the LHG mechanism in the context of
an observation by Adler on quantum-gravity induced
de-coherence effects.
Section 5 is devoted to conclusion.


\vspace*{1pt}\textlineskip	
\section{Brief review of AJY flux equalization}	
\vspace*{-0.5pt}
\noindent
Without CP violation, the three-flavor neutrino oscillation
framework carries five phenomenological parameters. These are the two
mass-squared differences, $\Delta m^{2}_{32}$ and $\Delta m^{2}_{21}$,
and the three mixing angles:

\def\ct{c_\theta}
\def\st{s_\theta}
\def\cb{c_\beta}
\def\sb{s_\beta}
\def\cp{c_\psi}
\def\sp{s_\psi}

\begin{equation}
U(\theta,\beta,\psi)=
\bordermatrix{
& 1 & 2 & 3 \cr
e   &\ct\cb & \st\cb & \sb \cr
\mu &-\ct\sb\sp - \st\cp & \ct\cp-\st\sb\sp & \cb\sp \cr
\tau    &    -\ct\sb\cp+\st\sp & -\st\sb\cp-\ct\sp & \cb\cp}
\end{equation}
The columns of the mixing matrix $U$ are numbered by the
mass eigenstates, $\jmath=1,2,3$, while the
rows are enumerated by the flavors, $\ell=e,\mu,\tau$. Here,
we have introduced the usual abbreviations: $c_x=\cos(x)$, and $s_x=\sin(x)$.

For a phenomenological study, the essential question is what are
the parameters of the neutrino oscillations and what information
may be extracted from them about particle physics, and astrophysical
and cosmological processes/sources. New flavor-sensitive
detectors with a collection area exceeding $1$ km$^2$ shall provide us
valuable information about the high-energy cosmic neutrino flux. This flux
carries important information about the conventional processes
of AGNs and GRBs, but it may  also serve as a probe of certain quantum
gravity effects and explore possible
violations of the equivalence
principle.\cite{qg1,qg2,qg3,qg4,qg5,qg6,qg7,qg8}

For high energy neutrinos, $E \simge 10^6$ GeV, with sources in AGNs
and GRBs, the source-detector distance far exceeds the kinematically
induced oscillation lengths suggested by any of
the solar, atmospheric, and the LSND data. Under these circumstances
the AGN and GRB neutrino flux, $F^S$, at the source is roughly in the ratio:
\beq
F^S_e:F^S_\mu:F^S_\tau :: 1:2:0
\label{agn_grb}
\eeq
The oscillated flux, $F_\ell^D$, measured at terrestrial detectors,
becomes independent of the mass squared differences,
and is given by:\refcite{yasuda}
\beq
F^D_\ell = \sum_{\ell^\prime}
P_{\ell\ell^\prime} F^S_{\ell^\prime}\label{yasuda}
\eeq
with
\beq
P_{\ell\ell^\prime} = \sum_\jmath \vert U_{\ell \jmath}\vert^2
 \vert U_{\ell^\prime \jmath}\vert^2
\label{p}
\eeq
Using the solar, reactor, atmospheric, and the accelerator,
neutrino data AJY have made
a detailed numerical analysis. The result
is~\refcite{yasuda}\footnote{Also see,
Ref.~\refcite{ajy}}:
\beq
\mbox{AJY's numerical analysis:}\qquad
F^D_e:F^D_\mu:F^D_\tau :: 1:1:1
\label{num}
\eeq
Analytically,\cite{ajy} AJY show
that the above result follows from the data-dictated assumptions:
\beq
\vert U_{e3}\vert^2 &\ll& 1, \nonumber\\
\left\vert \vert U_{\mu\jmath}\vert^2 -
\vert U_{\tau\jmath}\vert^2\right\vert &\ll& 1, \quad \jmath=1,2,3.
\label{ajy_anal}
\eeq


\vspace*{1pt}\textlineskip	
\section{Quasi Bi-maximal origin of flux equalization and an
ambiguity theorem}	
\vspace*{-0.5pt}
\noindent
We now show that this result is in fact a direct consequence
of the quasi bi-maximal mixing inferred from the $L/E$-flatness of the
electron-like event ratio observed in the Super-Kamiokande data on
atmospheric neutrinos. Then, in the next section, we show
that the flux equalization is not a unique signature of the source flux
given by Eq.\ (\ref{agn_grb}).

It was argued in Refs.~\refcite{qbm1,qbm2} that the observed
 $L/E$-flatness of the
electron-like event ratio in the Super-Kamiokande data on
atmospheric neutrinos places severe {\em analytical\/} constraints on the
mixing matrix. Without reference to the solar neutrino deficit, or the
data on the LSND excess events, it was shown that these constraints
yield a quasi bi-maximal  mixing matrix.\footnote{The quasi
bi-maximal mixing reduces to the bi-maximal mixing for $\theta=\pi/4$.
Apart from
Refs.~\refcite{qbm1,qbm2}, other
early references on bi-maximal mixing are~\refcite{bm1,bm2,bm3}.}
This result is contained in Eq.\ (26) of Ref.~\refcite{qbm2}, and reads:

\beq
U=
\left(\matrix{\ct & \st & 0 \cr
         -\st/\sqrt{2} & \ct/\sqrt{2} & 1/\sqrt{2} \cr
         \st/\sqrt{2} & -\ct/\sqrt{2} & 1/\sqrt{2} \cr}
\right)\label{bm}
\eeq
The mixing matrix (\ref{bm}), when coupled with Eq.\ (\ref{p}),
yields:\footnote{An invertible quasi bi-maximal mixing matrix $U$,
Eq.\ (\ref{bm}), necessarily yields a $P$ matrix that is non-invertible.
This mathematical observation shall underlie the physical content
of the theorem to be presented below.}
\beq
P=
\left(
\matrix{\st^4+\ct^4 & \ct^2\st^2 & \ct^2\st^2 \cr
\ct^2\st^2 & \frac{1}{4}\left[1+\st^4+\ct^4\right] &
\frac{1}{4}\left[1+\st^4+\ct^4\right] \cr
\ct^2\st^2   & \frac{1}{4}\left[1+\st^4+\ct^4\right] &
\frac{1}{4}\left[1+\st^4+\ct^4\right]}
\right)
\eeq
Substituting the obtained P in Eq.\ (\ref{yasuda}) furnishes with the
prediction:
\beq
\mbox{Quasi Bi-maximal mixing:}\qquad F^D_e:F^D_\mu:F^D_\tau :: 1:1:1
\label{analytical}
\eeq
This is precisely the result (\ref{num}) which AJY obtained
based on a detailed numerical analysis.\cite{yasuda} On the analytical
side,\cite{ajy} the AJY constraints (\ref{ajy_anal}) are manifestly
satisfied by the quasi bi-maximal mixing matrix (\ref{bm}).

Clearly, the AGN/GRB related ${\cal F}^S$ satisfy this flux equalization
criterion with $a=2$. For supernovae explosions, $a\approx 1$. Once again,
one obtains the flux equalization. The early results of Learned and
Pakvasa\cite{lp}, and Weiler {\em et al.\/}\cite{tw},
are seen to follow as a special
case associated with $\theta=0$ and $a=2$.

The result (\ref{analytical}) is independent of the mixing angle
$\theta$ -- the angle relevant for the solar, or LSND, data (see
Refs.~\refcite{qbm1,qbm2}).
This implies that the high
energy cosmic neutrino flux is robust in that it does not depend
on the (vacuum) mixing angle obtained from the solar neutrino anomaly,
or from the LSND data.
This robustness can be exploited to systematically study
other possible
significant sources of neutrino flux, especially those which may arise from
sources other than the decay of charged pions. The latter component
of the neutrino flux may appear as a departure from the
evenly proportioned flux of the three neutrino flavors discussed here.
The departures may also serve as a probe of certain quantum
gravity effects and possible
violations of the equivalence
principle.\cite{qg1,qg2,qg3,qg4,qg5,qg6,qg7,qg8}
However, we now emphasize that detecting a flux (\ref{analytical}) does not
necessarily imply the source flux to be (\ref{agn_grb}).

In interpreting any deviations from the result (\ref{analytical}), one must
be careful to note the following ambiguity theorem. Let
\beq
{\cal F}^S \equiv F^S_e: F^S_\mu: F^S_\tau :: 1:a:2-a, \quad 0\le a
\le 2
\label{s}
\eeq
Then, under the already stated framework, the quasi bi-maximal mixing
has the effect
\beq
{\cal F}^S \rightarrow {\cal F}^D
\eeq
where
\beq
{\cal F}^D \equiv F^D_e: F^D_\mu: F^D_\tau :: 1:1:1
\label{d}
\eeq
The proof is straight forward.

From an aesthetic point of view, a view which is also consistent
with the existing data, the quasi bi-maximal mixing is
a strong candidate to emerge as the unitary matrix behind the neutrino
oscillations. The widely discussed
bi-maximal mixing,\cite{qbm1,qbm2,bm1,bm2,bm3,wd} as already
noted, is a special case of the quasi bi-maximal mixing. In this special case
one may introduce a unitarity-preserving deformation of the bi-maximality,
and constrain it by the existing data as follows:
\beq
U^\prime=
\left(\matrix{\cb/\sqrt{2} & \cb/\sqrt{2} & \sb \cr
               -(1+\sb)/2 & (1-\sb)/2 & \cb/\sqrt{2} \cr
               (1-\sb)/2 & -(1+\sb)/2 & \cb/\sqrt{2}}\right),\quad \beta\ll 1
\eeq
This deformed bi-maximal mixing transforms ${\cal F}^S$ given by
Eq. (\ref{s}) into
\beq
{\cal F'}^D \equiv {F^\prime}^D_e: {F^\prime}^D_\mu: {F^\prime}^D_\tau ::
1:1+(a-1)\sb^2:1+(1-a)\sb^2
\label{dd}
\eeq
and carries an essentially unique signature for the deformation parameter
$\beta$, and for the source flux parameter $a$ (associated with the class of
neutrino fluxes under consideration).

\vspace*{1pt}\textlineskip	
\section{Flux equalization in LHG mechanism}	
\vspace*{-0.5pt}
\noindent

In a pioneering paper,\cite{lhg} LHG have argued that
quantum-gravity induced EHNS de-coherence\cite{ehns} can
lead to a flux equalization for all flavors.
However, the length scale at
which such a de-coherence sets in has been a reason of
some discussion.\cite{Lisi,Adler} The essential argument
that Adler advances is, in essence, correct:\cite{Adler}
these are
effects of the quantum-gravity induced de-coherence on
relative phases, and not on the global phase, that are important.
However, Lisi {\it et al.}\cite{Lisi} counter, in a ``Note added''
to their work, that, at present, one needs to take a purely phenomenological
approach to study such effects.
Without addressing the specific objection of Adler, Lisi {\em et al.}
question the dimensional arguments of Adler. It is the apparent weakness
of Adler's dimensional argument  that saves in the end the suggestion
of Lisi {\em et al.} -- a suggestion we suspect may, in fact, be too
optimistic for atmospheric neutrinos, but may carry viability for
astrophysical neutrinos.

In the context of the argument advanced by Lisi {\it et al.\/}
we note that dimensional arguments can indeed break down
when not all the relevant dimensionless numbers are incorporated
in one's calculation, or are not known.
The latter observation finds support
in the circumstance that
despite the standard ``forty orders of magnitude argument''
gravitationally-induced phases
in the neutron interferometry were first observed.\cite{COW}
The dimensionless numbers that were missed in the
standard arguments were:
(a) The mass of the Earth divided by
mass of the neutron, a number that equals $3\times 10^{51}$, and (b)
The dimensions of the
interferometer arm compared with the de Broglie wavelength
of a thermal neutron (which yields another
dimensionless number of about $10^{10}$).\cite{dva_nature99}. Similar
circumstance arises in detection of neutrinos (despite exceedingly
small cross sections), and in the possibility of probing space-time
fluctuations via gravity wave detectors.\cite{gac_nature99}

For  atmospheric neutrinos, a subject of the study
contained in Lisi {\em et al.}'s work,
$E$ is of the order of a few
GeV,  and $L$ ranges from about $20$ km to about $1.3\times 10^4$ km.
Compared with the atmospheric neutrinos, for astrophysical
neutrinos one gains in $L$ by a minimum
factor of about $10^{15}$, while for MeV range astrophysical
neutrinos, $E$ decreases by roughly three orders of magnitude.
Combined, these numbers give a minimum net
gain in $L/E$ by a factor of  about $10^{18}$.\footnote{For higher $E$,
$L/E$ suffers a decrease from this value. However,
VEP\cite{qg8} and qVEP\cite{qg1} effects increase
[no co-relation is implied, and final effects are a complex web of
different effects].}
Furthermore, one cannot assume that the quantum-gravity
induced de-coherence is independent of the relevant gravitational
environment. Such a dimensionless number
on the surface of the Earth is, $ M_\oplus G/R_\oplus c^2$ (incorporating
$c$ explicitly, now) $=7\times 10^{-10}$. Its counterpart on the surface
of a neutron star is about $0.2$.
These are precisely these dimensionless numbers that yield the
so called $20\%$ gravitational-induced effect in the red-shift
of flavor oscillation clocks in Ref. ~\refcite{grf96} (see Refs.
~\refcite{prd98,kk98,Lipkin,lhr01} for further discussion).
In astrophysical environments with intense gravitational fields,
by naive dimensional arguments, we may
obtain another ten orders of magnitude in favor of the
quantum-gravity induced de-coherence in neutrino oscillations.

The original arguments that suggested that the quantum-gravity
induced de-coherence may be observable for atmospheric, or solar,
neutrinos thus appears too optimistic. However, the relevance
of quantum-gravity effects in neutrino oscillations in
the cosmic/astrophysical contexts cannot be easily ruled out.
If the length scale for the  quantum-gravity
induced de-coherence for astrophysical neutrinos is of the order
of a Mpc (MegaParsec), then independent of the MNS matrix,
the LHG effect would lead to
the flux equalization for the  astrophysical neutrinos.


\vspace*{1pt}\textlineskip	
\section{Conclusion}	
\vspace*{-0.5pt}
\noindent
The observed $L/E$ flatness of the electron-like event ratio in the
Super-Kamiokande atmospheric data strongly favors a quasi bi-maximal mixing
for neutrino oscillations. This quasi bi-maximal mixing contains one
unconstrained mixing angle, $\theta$. The angle $\theta$ can either
be used to accommodate the LSND excess events, or to explain to the
long-standing solar neutrino anomaly. For high energy cosmic neutrinos,
the Source-Detector distance far exceeds any of the relevant kinematically
induced oscillations lengths. When this information is coupled with the
Super-Kamiokande implied quasi bi-maximal mixing, characterized by the angle
$\theta$, we find that a whole range of neutrino fluxes,
${\cal F}^S$, defined in Eq.\ (\ref{s}), and characterized by $a$, oscillate to
equal fluxes of
$\nu_e$, $\nu_\mu$, and $\nu_\tau$. This result carries a remarkable
robustness in its $\theta$- and $a$- independence.

Observation of equal $\nu_e$, $\nu_\mu$, and $\nu_\tau$ fluxes from
AGN/GRBs, and supernovae explosions, can be used to establish if they
belong to flux class, ${\cal F}^S$, defined above. Deviations
of these  fluxes,  ${\cal F}^D$, as observed in terrestrial detectors,
from the ratio $1:1:1$ can become a robust measure of
the departure
of the source flux ratio from $1:a:2-a$. A detailed study of these
departures carries the seeds to discover new physics, and to characterize
cosmic neutrino sources. In particular, it is to be emphasized that $a$
remains unmeasurable, if the mixing is quasi bi-maximal (of which,
bi-maximal mixing is a special case). Furthermore,
the angle $\theta$, that, e.g., can be adjusted to resolve the solar
neutrino anomaly,
does not influence the expected flux equalization. Because the high-energy
cosmic neutrino flux as detected in terrestrial laboratories
is insensitive to the underlying mass-squared
differences, measurements on the flavor spectrum
of the high-energy cosmic neutrino flux can be used to probe a
whole range of  parameters associated with neutrino oscillations. Since each
of these parameters -- from those related to the
deformed bi-maximal mixing,  to those that characterize a whole range of
quantum-gravity effects (including those violating the principle of
equivalence) -- is likely to yield a different
signature, high-energy cosmic neutrinos provide a powerful
probe into new physics.

The ambiguity that has been discussion of this {\em Letter\/}
is further compounded if the length scale for the  quantum-gravity
induced de-coherence for astrophysical neutrinos turn out to be
of the order of a  Mpc (MegaParsec). Then, independent of the MNS matrix,
the LHG effect would lead to
the flux equalization for the astrophysical neutrinos.


\nonumsection{Acknowledgments}
\noindent
I  thank  Sandip Pakvasa and Osamu Yasuda for remarks on
the first draft of this {\em Letter}. I am also thankful to
Eligio Lisi for a useful private communication.

\nonumsection{Note Added}

After this work was accepted for publication, Ker\"anen brought to my
attention his work where he and his colleagues\cite{bkm}
investigate the problem studied in this
{\em Letter\/} in the context of a two-doublet structure of
the 4-neutrino mixing. In a future study,
I intend to look at their scenario with
a special emphasis on the $L/E$ flatness of the e-like event ratio
seen in the atmopsheric neutrino oscillation data, and see to what extent
the ambiguity reported here survives in the extended frameworks of
neutrino oscillations.

Learned and Pakvasa (LP)\cite{lp} --  in a work predating  the AJY
result\cite{ajy} -- had also observed that a siginficant parameter space
of the $3\times 3$ neutrino oscillation framework contains the
flux equalization of different flavors. The LP study was based on
a whole range of allowed values of the three mixing angles as
deciphered from  then-existing data on the solar and atmospheric
neutrinos. Here, we have traced back the origin of the LP-AJY result to
the bi-maximal mixing as implied by the $L/E$ flatness of the e-like
event ratio observed at Super Kamiokande for atmospheric neutrinos,
and, in addition,  have brought in a new element that a similar ambiguity
in tracing back the source flux is introduced by certain quantum
gravity effects.


\nonumsection{References}

\end{document}